\documentstyle[twocolumn,aps,prl,epsfig]{revtex}

\newcommand{\ybco}{$\mathrm{YBa_{2}Cu_{3}O_{7-\delta}}$}

\newcommand{\loglog}{$\log(voltage)$ vs. $\log(current)$}

\begin{document}

\draft
\author{D. R. Strachan~\cite{strachan}, M. C. Sullivan, P. Fournier,
S. P. Pai, T. Venkatesan and C. J. Lobb}
\address{Center for Superconductivity Research, University of Maryland,
College Park, MD 20742}
\title{Is there a vortex-glass transition in high-temperature superconductors?}
\date{\today}

\twocolumn[\hsize\textwidth\columnwidth\hsize\csname@twocolumnfalse\endcsname

\maketitle

\begin{abstract}
We show that DC voltage versus current measurements 
of a \ybco\ micro-bridge in a magnetic field can be collapsed onto
scaling functions proposed by Fisher, Fisher, and Huse,
as is widely reported in the literature. We find, however,
that good data collapse is achieved for a wide range of critical
exponents and temperatures.
These results strongly suggest that agreement with scaling
alone does not prove the existence of a phase transition.
We propose a criterion to determine if the data
collapse is valid, and thus if a phase transition
occurs. To our knowledge, none of the
data reported in the literature meet our criterion.
 
\end{abstract}



]





One of the more remarkable consequences
of research on high-temperature superconductors
is a new picture of the normal to superconducting
transition in a magnetic field. While the mean-field
Ginzburg-Landau description had been viewed as adequate
for low-temperature superconductors, a consensus
has emerged~\cite{bib1,bib2,bib3,bib4,bib5}
that a vortex-glass transition occurs in the high-temperature
superconductors. The strongest evidence has come from
$voltage$ vs. $current$ ($I-V$) measurements~\cite{bib6}
where data can be collapsed onto
two scaling functions, as proposed by
Fisher, Fisher, and Huse (FFH)~\cite{bib7}.

Despite a strong consensus that this data collapse
implies the transition, some workers
have suggested that the apparent agreement with
scaling is misleading~\cite{bib8a,bib8b}.
Some simulations of $I-V$ curves which are
based on models without a phase
transition have also been shown to 
collapse onto scaling functions within 
limited voltage ranges. 

It has been countered~\cite{bib4} that
the simulations invoke highly unphysical
parameters in order to obtain
``scalable'' data which resemble actual measurements.
Moreover, the critical exponents found from the simulated
data~\cite{bib8a,bib8b} differ drastically from the ones obtained
experimentally.
Fueling the debate still further have
been some recent attempts at measuring $I-V$
curves over larger voltage ranges, where 
critical exponents have been found~\cite{bib5,bib10} 
which approach those resulting
from the controversial simulations~\cite{bib8a,bib8b}.

Furthermore, it has
recently been proposed~\cite{bib9} 
that a true phase transition,
such as the vortex-glass transition, does
not actually occur. This ``window-glass''
scenario is more like a conventional glass transition, where
the dynamics slow down considerably over a small temperature range, 
but correlation length scales do not strictly diverge. If
a superconductor were to behave in this way, the linear resistance could
rapidly decrease upon lowering temperature, but would not 
become zero. A small but non-vanishing linear resistance
has also been predicted by theoretical studies that 
incorporate screening~\cite{bib9b,bib9c}.

Granting all this, we do not see how this issue
can be resolved through the use of 
simulated $I-V$ curves as in Refs.\ \onlinecite{bib8a} and
\onlinecite{bib8b}. 
Simply showing that  
simulated data from a model without a transition
scales does not demonstrate 
that the measurements scale for the same reason.
The scaling interpretation of 
measurements may still be valid.

What is needed is an unambiguous signature in the data
that can be used to make a valid claim for a transition.
The purpose of this letter is to propose such a criterion and
to show that it is necessary. 
We propose 
that a criterion for determining whether
data supports a nonarbitrary choice of the critical 
parameters is that
\loglog\ isotherms equally distanced (\emph{ie.} with equal
$\left| (T - T_{g}) / T_{g} \right|$) from the critical temperature,
$T_g$, must have opposite concavities
at the \emph{same} applied currents. 
This contrasts with previously published data where opposite concavity
is seen, but \emph{not} at the same applied currents. 

\begin{figure}
\epsfig{file=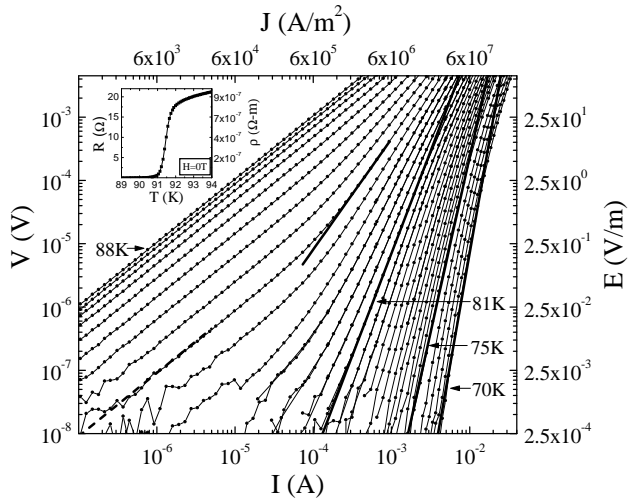,clip=,silent=,width=3.3in}
\caption{$I-V$ isotherms for a $2200 \mathrm{\AA}$ thick
\protect\ybco\ film in 4 tesla. The 
dashed line has a slope of one, and represents ohmic behavior,
while the solid lines are linear fits to the non-ohmic
power law-like  regions
of various isotherms. The inset shows an R(T) in ambient field.}
\label{fig:IV}
\end{figure}

We begin by showing
the difficulties in 
experimentally demonstrating that a
true phase transition exists in a superconductor. We start with a typical 
scaling analysis of
$I-V$ data taken on a
$\mathrm{2200 \AA}$ thick \ybco\ film laser
ablated onto an STO substrate.
This sort of film has previously
been shown to have transport characteristics
which agree with a scaling analysis
in externally applied magnetic fields
of about 4 tesla~\cite{bib1,bib2,bib3,bib4,bib5}. 
The high quality of our film was
verified with x-ray diffraction peaks
of predominately c-axis orientation, 
from an AC susceptibility measurement
transition width of 0.2K in 
zero magnetic field, and through zero
field R(T) (inset of Fig.\ \ref{fig:IV}) which shows
a $T_{C} \approx 91.5K$ and a transition width 
of about 0.5K. The film was photo-lithographically patterned into
a four terminal bridge $8 \mu m$ wide by $40 \mu m$ long
and etched using a dilute solution of phosphoric acid with no
noticeable degradation of R(T).

Figure\ \ref{fig:IV} shows $I-V$ measurements taken on our film in
a perpendicular magnetic field of 4 tesla. 
The scaling analysis requires that~\cite{bib7} 
\begin{equation}
V \xi^{2+z-D} / I= \chi_{\pm} ( {I\xi^{D-1}}/{T} ) \label{eq:scaling}
\end{equation}
where the dimensionality, $D$, is 3, $z$ is the
dynamic critical exponent, $\xi$ is the glass 
correlation length which is expected to behave as
$\left|(T-T_{g}) / T_{g} \right|^{-\nu}$, $\nu$
is the correlation-length exponent, and $\chi_{\pm}$ are
the scaling functions for above and below the
glass transition temperature $T_g$.

The parameters of Eq.\ (\ref{eq:scaling}) are found from
experimental data in the standard way. First, only 
those $I-V$ isotherms above $T_g$ should show 
low-current ohmic tails, where ohmic behavior is
represented in Fig.\ \ref{fig:IV} by the dashed line on the lower left
with a slope of 1. At higher currents the isotherms
are non-ohmic, and it is 
typically presumed that they
cross over to power law behavior (\emph{ie.}
straight lines on $log-log$ plots with slope greater than 1). 

The thick solid line 
at 81K is a power-law fit to the isotherm which separates
those with low-current ohmic tails from the ones without.
This is conventionally designated as $T_g$ 
and the slope of the fitted line on this plot
gives the dynamic exponent of $z=5.46$, since
$V \propto I^{(z+1)/2}$ is expected at $T_g$
from Eq.\ (\ref{eq:scaling}). 
The static exponent can be found from
the low-current ohmic tails, $R_L$, which are expected
to behave as
\begin{equation}
R_{L} \propto ({T/T_g}-1)^{\nu(z-1)}. \label{eq:RL}
\end{equation}

\begin{figure}
\epsfig{file=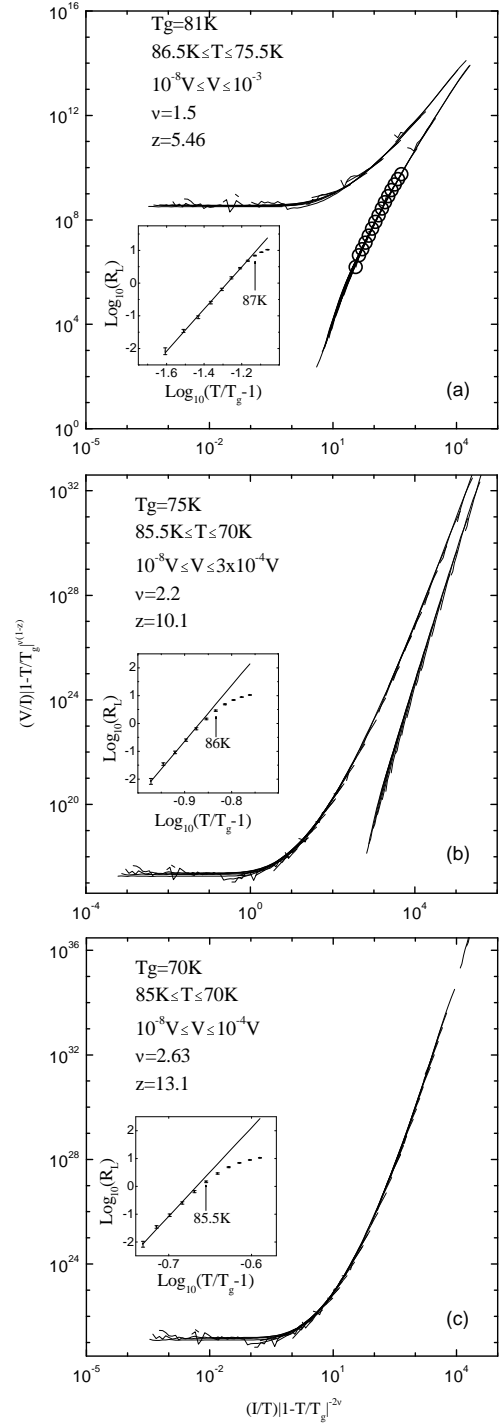,clip=,height=7.5in}
\caption{Data collapses of the $I-V$ curves
of Fig.\ \ref{fig:IV} using various critical parameters
with the experimental windows for each one denoted. The conventional
analysis is the one shown in (a). The insets show the $R_L$
plotted according to Eq.\ (\ref{eq:RL})}
\label{fig:comparison_final}
\end{figure}

Using the $z$ and $T_g$
found above we plot the $log_{10}$ of both sides 
of Eq.\ (\ref{eq:RL}) in the inset
of Fig.\ \ref{fig:comparison_final}(a). Since scaling predicts that
this plot should be a straight line, deviations from this at about 
87K determine the extent of the critical 
region, which is $\pm \mathrm{5.5K}$ from $T_g$. Only data
lying within this temperature window will be used to test 
Eq.\ (\ref{eq:scaling}), below.


Data at high currents are also excluded because
free flux flow occurs which is not described by scaling~\cite{bib1,bib5}.
A cut off is conventionally set to the voltage 
where the critical isotherm begins to deviate towards 
ohmic behavior, which is seen as a slight
decrease in slope at about $10^{-3}V$ in Fig.\ \ref{fig:IV}. 
Plotting all the data in Fig.\ \ref{fig:IV} below 
this voltage and within the 11K range about 81K,  
a conventional data collapse is clearly demonstrated in 
Fig.\ \ref{fig:comparison_final}(a) with
the critical exponents in good agreement with those reported elsewhere,
$z=5.46$ and $\nu = 1.5$.

\begin{figure}
\epsfig{file=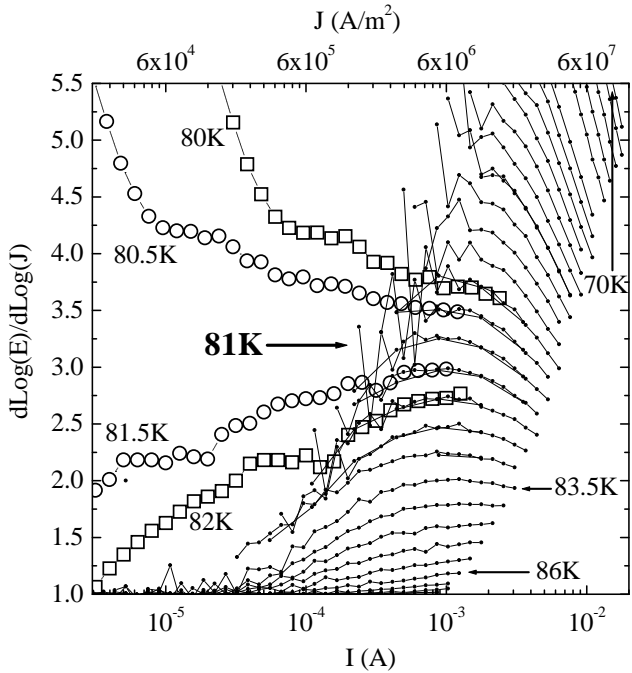,clip=,width=3.3in}
\caption{$dlog(V)/dlog(I)$ isotherms from
Fig.\ \ref{fig:IV} are plotted as small solid dots.
The open squares are extrapolated data which are 1K from
$T_g$ while the open circles are ones 0.5K from $T_g$.
Both pairs of extrapolated data were 
extracted from the collapse shown in Fig.\ \ref{fig:comparison_final}(a).}
\label{fig:dlog_final}
\end{figure}

There is a serious problem with the analysis
outlined above. Following Repaci \emph{et al.}~\cite{bib_repaci}, 
we demonstrate this by plotting the derivatives
of the \loglog\ isotherms, which are shown in Fig.\ \ref{fig:dlog_final}
as small solid dots. The predicted power-law
$I-V$ curve at $T_g$ should correspond to a horizontal
line in Fig.\ \ref{fig:dlog_final}, with a value
of $(z+1)/2$. The data, however, \emph{peaks} at about
$7 \times 10^{-4} \mathrm{A}$. In fact,
all isotherms seem to have a maximum slope at about this 
current with ohmic tails developing to the left of the peaks. 
Apparently, the
only difference between the isotherms above
and below the 
conventionally determined $T_g$ is
that the ones at lower temperatures are truncated 
due to the resolution limit of the experiment before they
decrease in slope towards ohmic behavior. This truncation
is evident in these derivative plots where 
the data at lower currents and voltages become noisier.

Since the conventionally chosen
critical isotherm does not show any signs of unique
power law behavior, we now ask
whether it is possible to choose other values for $T_g$
which also lead to data collapse. 
To do this we note that non-ohmic 
power law-like behavior can be fit to the lower temperature
isotherms over at least 4 decades of voltage
data. We have demonstrated this in Fig.\ \ref{fig:IV}
for the isotherms at 75K and 70K. By repeating the 
scaling analysis assuming that these two temperatures
are the critical isotherms we can again obtain excellent
data collapses as is demonstrated in Figs.\ \ref{fig:comparison_final}(b)
and \ref{fig:comparison_final}(c).


When we lower the defined $T_g$ by 6K in going from 
Fig.\ \ref{fig:comparison_final}(a) to Fig.\ \ref{fig:comparison_final}(b)
we need only readjust $z$ and $\nu$ in order to
maintain successful data collapse with an even larger
critical region (greater than 15K). As shown in 
Fig.\ \ref{fig:comparison_final}(c), $T_g$ can even be
defined as the lowest temperature of 
our measurement, where $\chi_{-}$ is not shown because
there is no data below 70K.

Now that we have demonstrated that a data collapse 
does not uniquely determine the critical parameters, we 
propose a criterion for uniquely determining
$T_g$, and thus $\nu$ and $z$.

Such a signature is suggested by
the scaling functions found from the conventional
vortex-glass analysis. To see this
it is important to note that each
isotherm in Fig.\ \ref{fig:IV}
collapses onto only small portions
of the scaling functions of Fig.\ \ref{fig:comparison_final}.
We demonstrate this by plotting only the isotherm at 79K
as open circles in Fig.\ \ref{fig:comparison_final}(a).
In the low current direction of the 
collapses the isotherms are cut off
by the voltage sensitivity floor of the experiment.

We can, however, predict how data at lower
voltages would behave if a data collapse
is assumed to represent a real transition.
We do this by using the 
temperature for the desired isotherm, a current,
and the values of the parameters $\nu$ and $T_g$
used in the collapse.  
This information determines a position along
the horizontal axis of Fig.\ \ref{fig:comparison_final}(a). 
Now, by using the vertical axis value of the scaling
function along with the $z$ exponent we can solve for
the predicted voltage. 


\begin{figure}
\epsfig{file=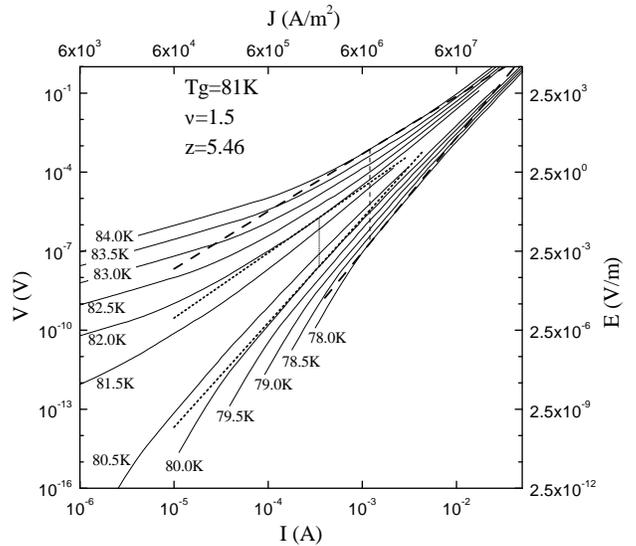,clip=,width=3.3in}
\caption{Extrapolated $I-V$ data from the collapse of
Fig.\ \ref{fig:comparison_final}(a). Dashed and
dotted lines are tangents drawn to the isotherms in
order to demonstrate opposite concavity at common applied currents.}
\label{fig:sim}
\end{figure}


The results of the extrapolation are shown in 
Fig.\ \ref{fig:sim}, 
with the $I-V$ curves displaying 
a property not seen in the measured data. 
For isotherms at equal temperatures away from $T_g$ 
(\emph{ie.} equal $\left| (T - T_{g}) / T_{g} \right|$),
opposite concavities are clearly evident
\emph{at the same current level}. We 
demonstrate this in Fig.\ \ref{fig:sim}
with vertical lines which represent 
constant currents drawn between two different pairs of isotherms.
Tangent lines to these isotherms at the intersections
clearly show that both pairs have opposite
concavity at the same applied currents.
The reader can verify that this signature would restrict
the assignment of $T_g$ to within $\pm \mathrm{0.5K}$ 
of 81K whether the resolution
of the experiment would 
be at $10^{-16}\mathrm{V}$ or $10^{-10}\mathrm{V}$~\cite{bib11}
by covering the extrapolated data at low voltages.

We demonstrate the striking contrast between this extrapolated
data and the real data by plotting the extrapolated
ones as open squares and circles on the derivative plot
of Fig.\ \ref{fig:dlog_final}. Note that the actual data
curves in Fig.\ \ref{fig:dlog_final} are all qualitatively
the same. It is only in the extrapolated region that curves
with equal $\left| (T - T_{g}) / T_{g} \right|$ 
show opposite concavity at the same applied currents.

Since opposite concavity could only envelope
a single temperature at a specified current, the
measurement of this signature would preclude
the arbitrary choice in $T_g$ that a conventional 
scaling analysis permits.  Thus, to allow for an 
adequate determination of
$T_g$, it is necessary to see negative 
concavity for an isotherm below $T_g$,  
while one above and with equal 
$\left| (T - T_{g}) / T_{g} \right|$
has a positive concavity~\cite{bib12}.  
Since the criterion is not satisfied
by our data nor the ones we know of in the literature, 
we argue that true evidence for a vortex-glass phase transition 
has not yet been demonstrated through $I-V$ characteristics.

In conclusion, we have found that a data 
collapse is not sufficient evidence for 
a vortex-glass transition since the critical parameters can be chosen
in various combinations, while maintaining agreement with scaling.
Furthermore, we have shown that our $I-V$ data 
plus those in the literature are not consistent with
a true phase transition because the 
experimentally determined scaling functions
predict a signature of the transition never seen, yet which should
be experimentally accessible. Measurement of this signature
would prevent the arbitrary choice of critical parameters
permitted by the conventional analysis.  Furthermore, this 
signature can be used as a criterion 
to judge future $I-V$ data in order to help settle
the controversy surrounding critical phenomena in the
high-temperature superconductors.

The authors would like to thank
S. M. Anlage, A. Biswas, Georg Breunig, R. C. Budhani, Zhi-yun Chen, 
R. L. Greene, P. Minnhagen, R. S. Newrock, A. P. Nielsen, and A. Schwartz 
for useful discussions on this work. We would also like to acknowledge the 
support of the National Science Foundation
through Grant No. DMR-9732800.






\begin{references}
\bibitem{bib11}
This is an experimentally accessible voltage resolution. See for example Ref.\
  \onlinecite{bib3}.
\bibitem{bib12}
We require that the relevant temperature scale of the transition, $\left| (T -
  T_{g}) / T_{g} \right|$, be equal for the two isotherms since both must be in
  the critical region, which excludes the possibility of comparing critical to
  non-critical behavior.
\end{references}

%
%

%
%

\end{document}